\documentclass[%
 reprint,
 amsmath,amssymb,
 aps,
]{revtex4-1}

\usepackage{graphicx}
\usepackage{dcolumn}
\usepackage{bm}

\begin{document}


\title{Through-Bottle Whisky Sensing and Classification using Raman Spectroscopy\\in an Axicon-Based Backscattering Configuration}

\author{Holly Fleming$^1$}
\author{Mingzhou Chen$^1$}%
\author{Graham D. Bruce$^1$}
\author{Kishan Dholakia$^{1,2}$}
\email{kd1@st-andrews.ac.uk}
\affiliation{$^1$SUPA, School of Physics and Astronomy, University of St Andrews, Fife, KY16 9SS, United Kingdom}%
\affiliation{$^2$Department of Physics, College of Science, Yonsei University, Seoul 03722, South Korea}%

\date{\today}

\begin{abstract}
Non-intrusive detection systems have the potential to characterise materials through various transparent glass and plastic containers. Food and drink adulteration is increasingly problematic, representing a serious health risk as well as an economic issue.  This is of particular concern for alcoholic spirits such as Scotch whisky which are often targeted for fraudulent activity. We have developed a Raman system with a novel geometry of excitation and collection, exploiting the beam propagation from an axicon lens resulting in an annular beam that transforms to a Bessel illumination within the sample. This facilitates the efficient acquisition of Raman signals from the alcoholic spirit contained inside the bottle, while avoiding the collection of auto-fluorescence signals generated by the bottle wall. Therefore, this technique provides a way of non-destructive and non-contact detection to precisely analyse the contents without the requirement to open the bottle. 
\end{abstract}

\maketitle


\section{\label{sec:intro}Introduction}

Raman spectroscopy is a powerful and popular label-free method to gain information at the molecular level. In particular, it has demonstrated itself as a reliable material identification technique in a wide range of applications including biomedicine\cite{baker2016fundamental,chen_use_2015,schlucker_immuno-raman_2006,jamieson_bioanalytical_2017, Baron2017}, cultural heritage\cite{conti_comparison_2015,realini_development_2017,conti2016portable}, and defence.\cite{ramirez-cedeno_remote_2010,moore2009portable,loeffen_spatially_2011} Typically, Raman spectroscopy instruments employ a backscattering configuration, where the Raman scattered photons are partially collected through the same optical path as the incident light.  While conventional Raman spectroscopy is a sensitive analytical technique used to produce unique chemical fingerprints of materials, it can be limited by sample volume and depth, and often swamped by signals from any material present in a particular optical path, before the sample plane itself. Of particular interest, is the ability to sense through sealed bottles, negating the requirement of opening the containers to spectrally examine the contents.  

The UK spirits market currently stands at around \pounds13.5 billion with whiskies and gins proving popular choices.\cite{alcoholmarketseg2019} In particular, Scotch whisky is a high value product often retailing at hundreds of pounds or US dollars, with the price usually increasing with the maturation age. In cases of suspected fraud or adulteration, \textit{in situ} sensing without the need to open the bottle would be preferable in order to retain the original value of the product. 

Whisky is a chemically complex mixture, comprising thousands of compounds which contribute to the colour, flavour and aroma of the beverage. Several studies have previously looked at a detailed analysis of the composition using optical spectroscopic methods such as Raman\cite{ashok_near_2011, kiefer_analysis_2017} and infrared (IR) spectroscopy\cite{sujka2018application}, or non-optical techniques such as nuclear magnetic resonance (NMR) and mass spectrometry (MS).\cite{kew_chemical_2017,kew2019analysis} However, many of these studies require direct analysis of the sample, and thus cannot be carried out \textit{in situ}. One of the major obstacles with through-bottle sensing in this case is that glass is a highly Raman active substance, often masking regions of interest within the bottle. Studies investigating sealed contents of bottles have primarily focused on the detection of specific food denaturants\cite{ellis_through-container_2017,ellis2019rapid}, or the detection of concealed hazardous substances, using alternative Raman configurations such as spatially offset Raman spectroscopy (SORS).\cite{ramirez-cedeno_remote_2010,moore2009portable,loeffen_spatially_2011,nicolson_through_2017}

Conventional Raman configurations operate with the optical path of the collection light directly overlapping with the excitation laser beam, i.e. being collinear and confocal, often using a microscope objective. This makes the efficiency of collection high. However, it suffers from more unwanted signals from surrounding materials that are present in the final spectrum. SORS provides a way to avoid this by purposely shifting the collection point away from the excitation focal point, usually in the order of a few millimetres.\cite{maher2010determination, nicolson_through_2017, Chen2017} In doing so, there is a significant reduction in the Raman signal intensity from the surface layers but also in the Raman signal intensity from the region of interest, consequently requiring a much longer integration time or higher laser powers. Alternatively, the signals can be improved by using inverse SORS (iSORS) \cite{matousek_subsurface_2005, matousek_deep_2007,khan2016inverse} where a ring-illumination is used, centred about the collection region. Typically, an annular beam is produced either by using the far-field beam profile formed by an axicon\cite{duocastella2012bessel}, or by employing a multi-core fibre bundle with the fibers arranged on annuli of different radii.\cite{matousek2006noninvasive} The iSORS configuration has the added advantage of increasing the illumination area, therefore enabling higher laser powers to be used.\cite{matousek_subsurface_2005,khan2016inverse} The key limitation of SORS or iSORS is that the excitation beam is focused at the surface or at a point inside the sample while the collection is at a distance away from the position with the strongest excitation (focus).\cite{matousek_inverse_2006} This may result in the original, very weak Raman signals being obscured as well as containing strong background noise from the excitation points.

In this paper, we introduce a geometry of excitation and collection which can be employed for through-bottle sensing, by improving the signals recovered from the sample contained within the bottle, while reducing the contribution of the dominant glass surface to the collected Raman signal. The key advantage of our approach is using an axicon-based geometry that circumvents the need for a microscope objective, simply requiring a single axicon lens to produce an annular beam. We make use of the Fourier relationship between the annular beam formed by the axicon and the focused Bessel beam on light propagation.\cite{Chen2013,Vaity2015} 
This makes it possible to form the annular beam on the bottle wall while having a focused Bessel beam with central maximum of intensity inside the liquid contents. The back scattered fluorescence and Raman signals from this Bessel beam are collected through the hollow area of the annular beam on the bottle wall, thus avoiding collection of the Raman signals excited from the glass wall. Using this selective epi-geometry, we can exclude any undesired signals from the glass bottle with minimal loss of efficiency of collection from the liquid inside. 

\section{Material and Methods}
\paragraph{Axicon-Based Optical Setup\\}
\begin{figure}[b]
\centering
\includegraphics[width=0.9\columnwidth]{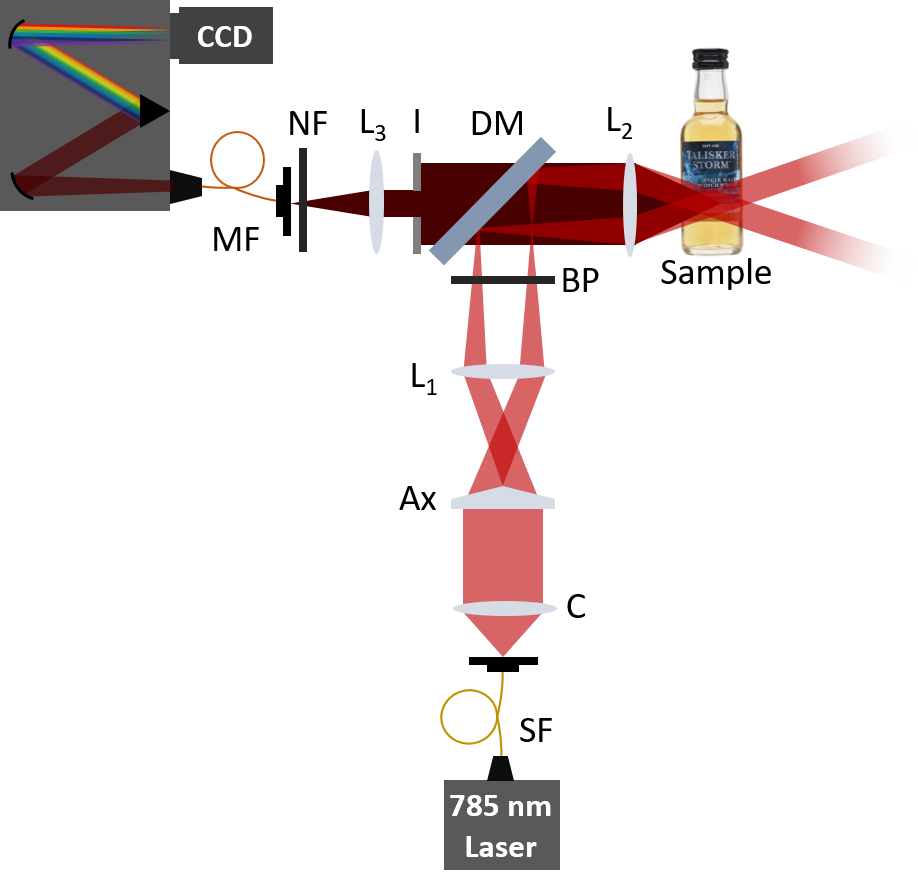}
\caption{Schematic illustration of the optical setup for through-bottle sensing using a backscattering configuration. SF: single-mode fibre; C: collimating lens; Ax: axicon lens (Thorlabs AX255-B); BP: laser line filter (Semrock LL01-785); DM: dichroic mirror (Semrock LPD02-785RU); I: iris; NF: notch filter (Semrock NF03-785E); MF: multi-mode fibre (200 $\mu$m core diameter); L\textsubscript{1}-L\textsubscript{3}: lenses.}
\label{scheme}
\end{figure}
The measurements were performed using the setup shown schematically in Fig.~\ref{scheme}. This is an optical illumination and collection system which was based on a similar arrangement as standard for epi-fluorescence measurements. 
A tunable Ti:Sapphire laser (Spectra-Physics 3900s, 785nm, maximum power 1W) is used as the Raman excitation source. It delivers 100 mW laser power into the sample via a line filter (LL01-785, Semrock, Rochester, NY, USA) which is used to appropriately filter the excitation beam. The collimated excitation beam passes through an axicon lens, Ax (Thorlabs AX255-B, 5$^{\circ}$ apex angle), which forms a Bessel beam at the back focal plane of the lens L$_1$. Lens L$_1$ performs a Fourier transform, converting the Bessel beam into a annular beam at the front focal plane of L$_1$. 
The annular beam is reflected by a dichroic mirror (LPD02-785RU, Semrock, Rochester, NY, USA) and then refocused by another lens L$_2$ into the sample. Here, a Fourier Transform by L$_2$ reverts the annular beam back into a Bessel beam at the focal plane, which is within the bottle (the region from which we wish to collect the Raman signals). Lenses L$_1$ and L$_2$ form a standard 4-f system to image the Bessel beam from the axicon to the sample plane. Comparing this with the traditional confocal Raman or SORS configurations, this axicon-based geometry provides a hollow area inside the ring beam before the final Bessel beam is generated inside the sample. The emitted Raman excitation from the Bessel beam central maximum is collected through the hollow area selectively using an iris (I). As a annular beam forms on the sample surface (e.g. glass bottle wall), the emitted Raman photons from the surface can be excluded from the collection path by appropriately adjusting the aperture size of the iris. Analogously with other Raman systems, Rayleigh scattered photons are removed by the dichroic mirror (DM) and a notch filter NF (NF03-785E, Semrock, Rochester, NY, USA). The Raman excited light passes through this mirror and filter and is collected by a spectrometer (a monochromator (Shamrock SR-303i, Andor Technology) with a 400 lines/mm grating, blazed at 850 nm, and a deep depletion, back illuminated and thermoelectrically cooled CCD camera (Newton, Andor Technology)).

\paragraph{SORS Optical Setup\\}
Using the same 785 nm excitation source as the axicon-based optical setup, 100 mW laser power is delivered into the sample via a line filter (LL01-785, Semrock, Rochester, NY, USA) which is used to appropriately filter the excitation beam. The collimated excitation beam is passed through a lens, with the focal point matching the location of the L$_2$ focal point, at an offset distance of 0 mm. 
The Raman scattered light is collected through L$_2$, following the same optical path as described in Fig.~\ref{scheme}, without the inclusion of an iris in the setup. A linear translation stage was used to move the optical collection path in a perpendicular direction away from the excitation focal point, providing offset illumination and collection points.

\paragraph{Experimental Samples\\}
A total of 12 different commercially available spirit drinks were analysed, each in their original bottles. The branded spirit drinks were purchased from national retail outlets and included a range of vodka, whisky, and gin spirits.

\paragraph{System Characterisation\\}
Initial characterisation of the optical setup outlined in Fig.~\ref{scheme} was performed to characterise the profile of the beam along the optical axis. Beam profile images taken with a FLIR Flea3 USB3 Camera and analysed using ImageJ. 

\paragraph{Raman spectral measurements of alcoholic drinks\\}
For the assessment of sample depth vs reduction of the glass signal, the sample bottle was positioned so that the surface of the glass bottle was located at the focal point of L\textsubscript{2}. The sample was moved along the optical axis in 2 mm increments, towards L\textsubscript{2}, recording the Raman spectra after each movement.

The reduction in glass signal using the axicon-based system was compared to that of a SORS system, taking the ratios of peak intensities of the glass (1370 cm\textsuperscript{-1}) and EtOH (881 cm\textsuperscript{-1}) contributions. Offset distances for the axicon-based system were calculated by using the radius of the annular beam on the surface of the bottle. Comparatively, the offset distances for the SORS system were determined by the distance between the excitation and collection points. 

To compare our axicon-based system with a conventional Raman set-up, the axicon (Ax), lens 1 (L\textsubscript{1}), and iris (I) were removed from the system (Fig.~\ref{scheme}). The Raman spectra obtained were processed by normalising them with respect to the EtOH peak at 881 cm\textsuperscript{-1} over a spectral range of 800 -- 1800 cm\textsuperscript{-1}. The optical power onto the sample plane was measured to be 100 mW. Each sample was placed at a surface distance of of 2.5 cm from L\textsubscript{2}. 

Raman spectra were acquired using Solis (s) Software (Andor Technology). Reference spectra for laser calibration were acquired from polystyrene beads. Standard acquisitions were conducted at 785 nm, using 5 accumulations of 5 seconds laser exposure. 

\paragraph{Spectral processing\\}
The obtained Raman spectra were processed using Origin (OriginLab, 2018). To compare the glass to ethanol ratio, spectra were truncated and baselined to remove as much fluorescent background as possible. Following baseline subtraction, the spectra were then normalised with respect to the ethanol (EtOH) peak intensity at 881 cm\textsuperscript{-1}. The data was subsequently smoothed. Multivariate analysis (principal component analysis, PCA) was used to classify the spirits based on their spectral features. This was performed on pre-processed Raman data, employing standard algorithms.

\begin{figure}[b]
\centering
\includegraphics[width=0.9\columnwidth]{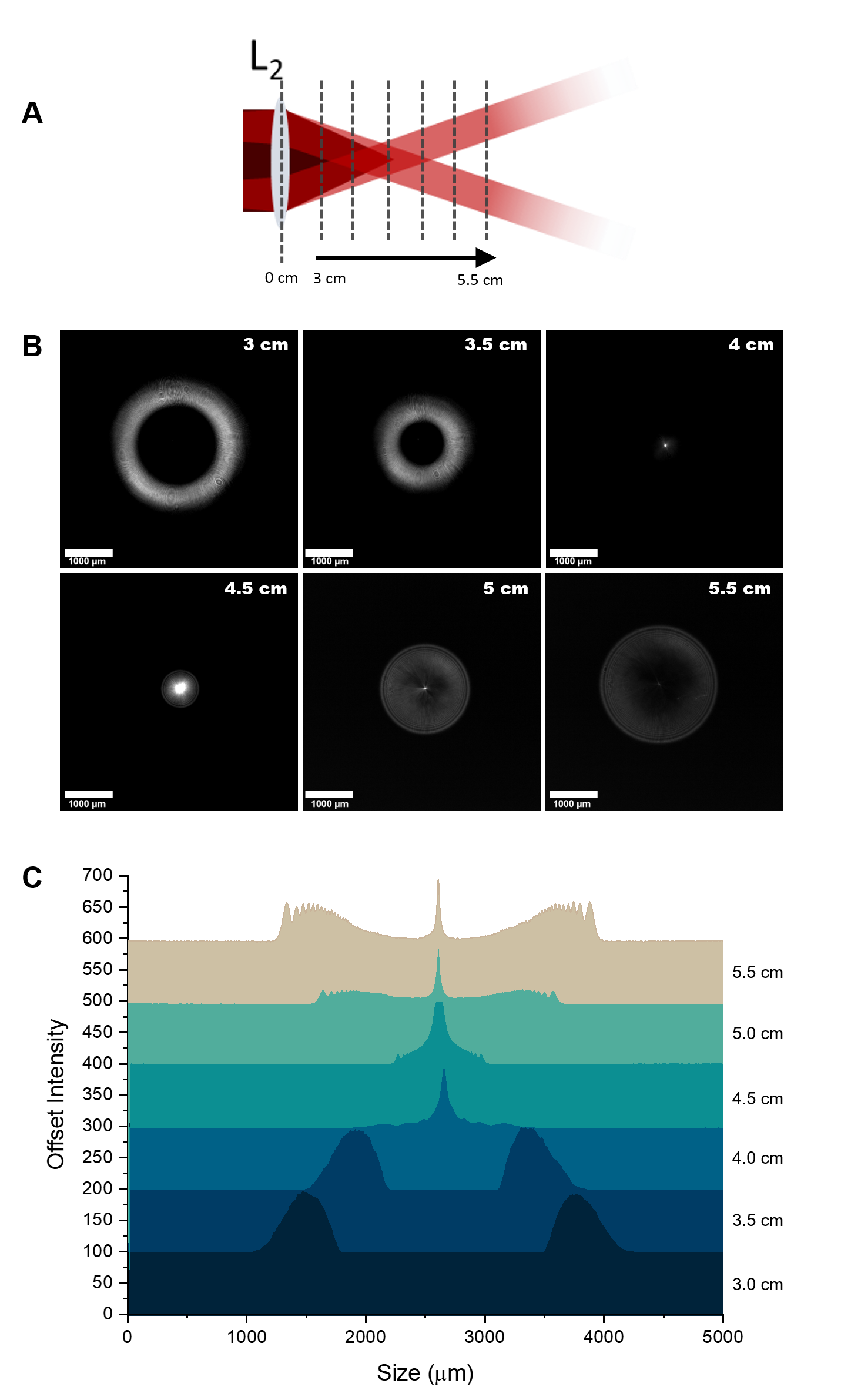}
\caption{Beam profile images. (A) Illustration of the position of the camera to take profile images. (B) beam profile images taken between 3--5.5 cm from L\textsubscript{2} showing the annular profile narrowing to form a Bessel beam (at 4 cm), with the beam subsequently expanding thereafter. Scale bar 1000 $\mu$m. (C) Intensity profiles across the beam images, showing the formation of a Bessel beam at the focal point of L\textsubscript{2}.}
\label{Beamprofiles}
\end{figure} 

\section{Results and Discussion}
\begin{figure}[bt]
\centering
\includegraphics[width=0.9\columnwidth]{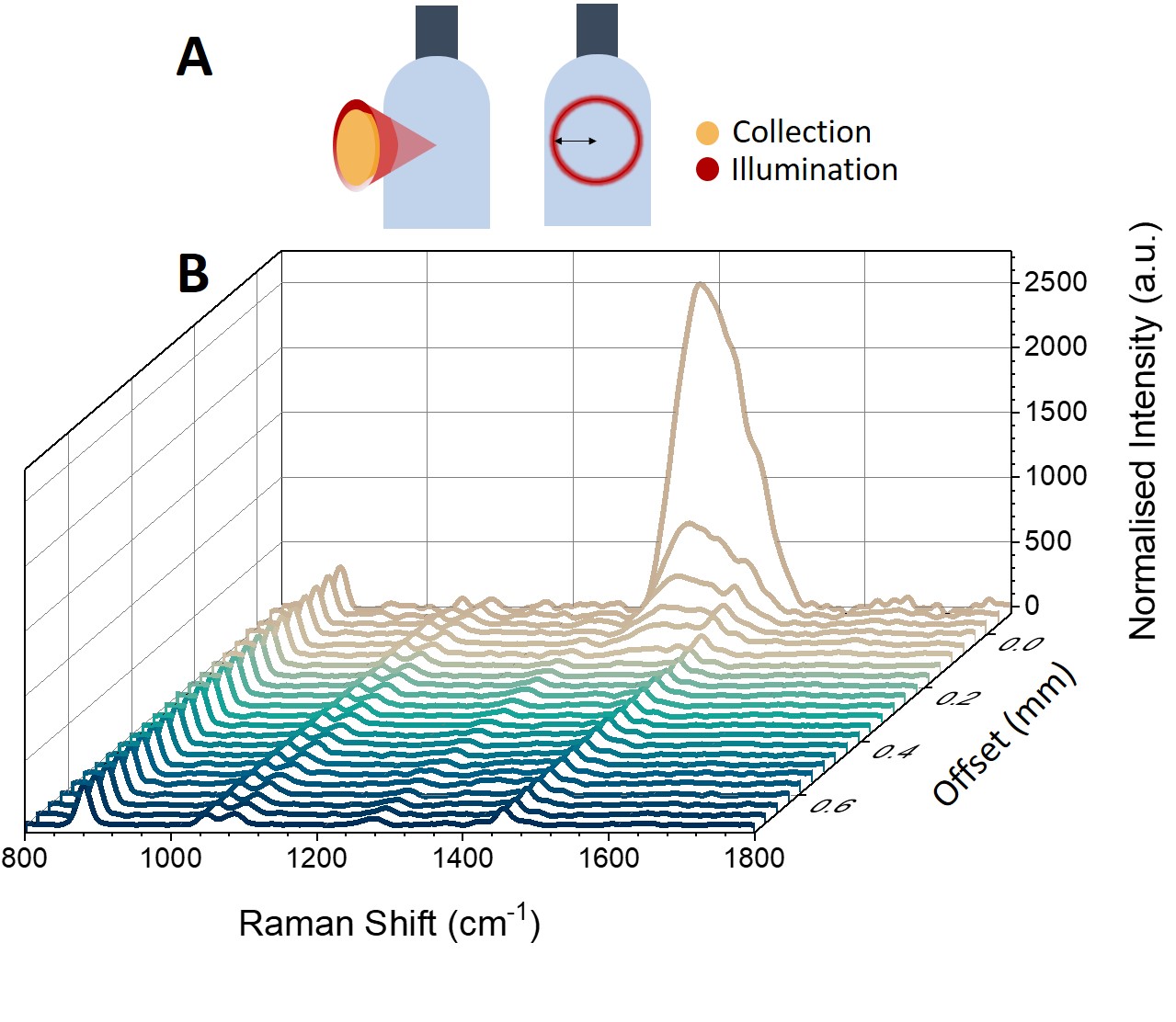}
\caption{Raman spectra through a glass bottle. (A) Illustration of the hollow cone-shaped light interacting with the drinks bottle and the collection of scattered light through the centre, and the resulting beam profile on the surface of the bottle. (B) Raman spectra of the sample bottle obtained by moving the sample in 2 mm intervals along the optical axis, demonstrating the suppression of the Raman signal due to glass (1370 cm\textsuperscript{-1}). Spectra have been normalised for clarity. Measurements were performed using a laser excitation wavelength of 785 nm, average laser power 100 mW, and 5 s integration time, five accumulations}
\label{Irisbottle}
\end{figure}

Spirit beverages were chosen due to their high economic value, of which a rapid, non-destructive, and non-invasive analytical procedure can prove valuable for monitoring purposes. A major component of distilled spirits is ethanol, with the content usually between 35-50$\%$ (alcohol by volume, ABV). This has been the focus of many previous studies\cite{kiefer_analysis_2017,ellis_through-container_2017,ellis2019rapid,ashok_near_2011}, and from our own measurements it is apparent that ethanol is a dominant spectral constituent (Fig.~\ref{Irisbottle}) showing the following strong vibrational peaks: C-C stretch at 881 cm\textsuperscript{-1}, C-O stretch at 1041 and 1080 cm\textsuperscript{-1}, and CH\textsubscript{2} wagging at 1450 cm\textsuperscript{-1}). Aside from ethanol, these spirits possess a complex chemical makeup, particularly in the case of Scotch whiskies which mature in wooden casks, adding to flavour, colour, and aroma, collectively known as congeners. Although low in volume, they present an analytical challenge as they often contribute to a highly fluorescent background. While fluorescent contributions pose a challenge in most Raman applications, here, the fluorescent background can be beneficial, as it can aid in the determination of whiskies and their composition.\cite{kiefer_analysis_2017,ashok_near_2011}

\paragraph{System Characterisation\\}
Our primary aim was to assess the ability of the system to measure the contents of a glass drinks bottle whilst suppressing the Raman signal afforded from the glass. The beam quality along the propagation is quite critical for both excitation and collection. Therefore the beam profile along the optical axis was firstly assessed. Images were taken at 5 mm intervals between 30 mm and 55 mm from lens L\textsubscript{2} (Fig.~\ref{Beamprofiles}). Fig.~\ref{Beamprofiles} B shows a hollow beam profile focusing to an apex, at a distance matching the focal length of L\textsubscript{2}. It is at this apex where the Bessel beam is formed. The apex of the annular light beam can be adjusted by exchanging L$_2$ with another lens of a different focal length. 

As shown in Fig.~\ref{Beamprofiles}, the ring size decreases after a distance of 4cm (the focal length of L$_2$), expanding again afterwards. A numerical simulation, using a split step Fourier beam propagation algorithm~\cite{Goodman2005}, qualitatively assessed the beam profile, using a glass slide of a similar thickness to mimic the bottle wall. This showed that the beam had undergone no significant deformation (numerical data not shown here) confirming the experimental validity of our approach.

\begin{figure}[!ht]
\centering
\includegraphics[width=0.9\columnwidth]{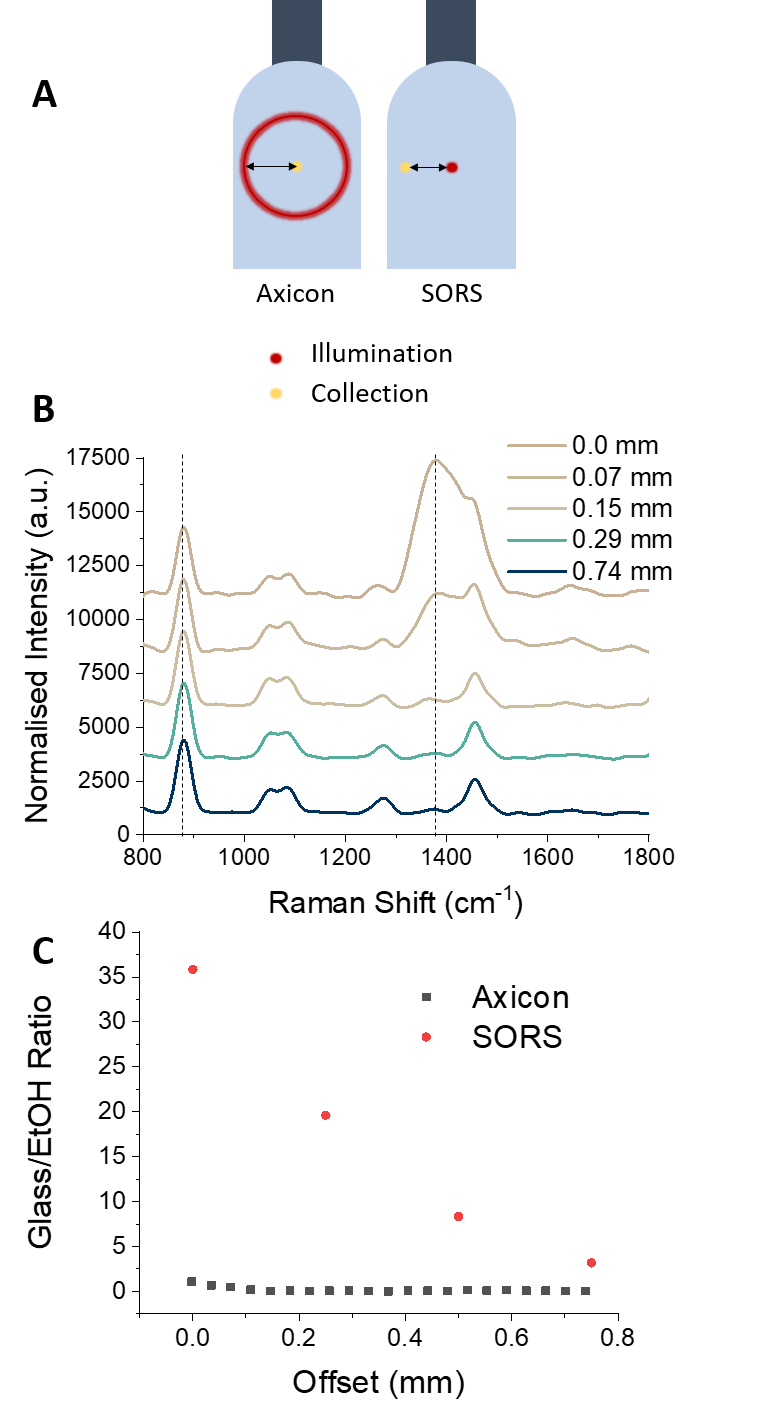}
\caption{Comparison between the axicon setup and SORS setup. (A) Offset distances are calculated by the radial distance at the surface of the bottle for the axicon setup, and by the distance between illumination and collection paths for the SORS setup. (B) The ratio of glass (1370 cm\textsuperscript{-1}) to ethanol (881 cm\textsuperscript{-1}) peaks was calculated (indicated by dashed lines). (C) Measurements were performed using a laser excitation wavelength of 785 nm, average laser power 100 mW, with five accumulations of each measurement using 5 s integration time.}
\label{AxVsSORS}
\end{figure}

The line profile of these images was then plotted showing the radius decreasing towards the focal point of L\textsubscript{2} (4 cm), and so confirming the annular form of the beam profile. 
The system presented here employs an excitation path that converges to a point, matching the focal length of the collection lens L$_2$. This is notably different to a typical iSORS configuration where the ring illumination is focused on the surface and continues to propagate outward after its formation, resulting in an increasing ring size throughout the sample.\cite{khan2016inverse,matousek_inverse_2006} 
Although the sample can be positioned so that the Bessel beam can be formed after the glass, the annular beam will unavoidably excite the glass to generate unwanted background signal. In order to eliminate the glass signal from entering the spectrometer, an iris was added to the setup (Fig.~\ref{scheme}) to avoid any collection from the region of the annular beam.

In addition, an advantage of our system is it requires only a single axicon element to generate the annular illumination beam, which could be introduced with minimal disruption into an existing conventional Raman optical system, making it a relatively simple and cost effective setup.

\begin{figure}[t]
\centering
\includegraphics[width=0.9\columnwidth]{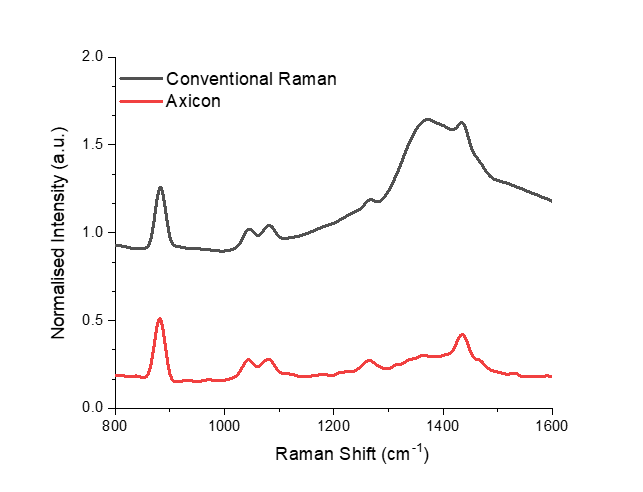}
\caption{Raman spectra of Talisker Storm through bottle using a conventional Raman backscattering configuration and the axicon setup (with sample at a 25 mm distance), showing suppression of the strong signal originating from the glass. Spectra have been normalised and truncated for clarity. Measurements were performed using a laser excitation wavelength of 785 nm, average laser power 100 mW, five accumulations using a 5 s integration time.}
\label{AxVsCr}
\end{figure}

\paragraph{Application\\}
Here, we demonstrate the axicon-based configuration for through-bottle sensing for the classification of whiskies and  other distilled spirits. 
Whiskies contain chemically complex mixtures, comprising a wide variety of organic compounds which contribute to the colour, aroma and taste of the beverages. 
During the maturation of the whisky, compounds from the casks diffuse into the solution, providing distinctive flavouring of a particular brand. 
It is these organic constituents (which provide a fluorescent component) combined with the Raman peaks afforded from ethanol that enable the differentiation of the beverages. 
However, unwanted signals originating from the glass bottle can significantly interfere with signals from the bottle's contents. 
As such, the spectra obtained from a glass bottle by employing a conventional backscattering Raman configuration are typically dominated by a large peak at 1370 cm\textsuperscript{-1}, arising from glass signal contributions (Fig.~\ref{AxVsCr}). 

\begin{figure}[t]
\centering
\includegraphics[width=0.9\columnwidth]{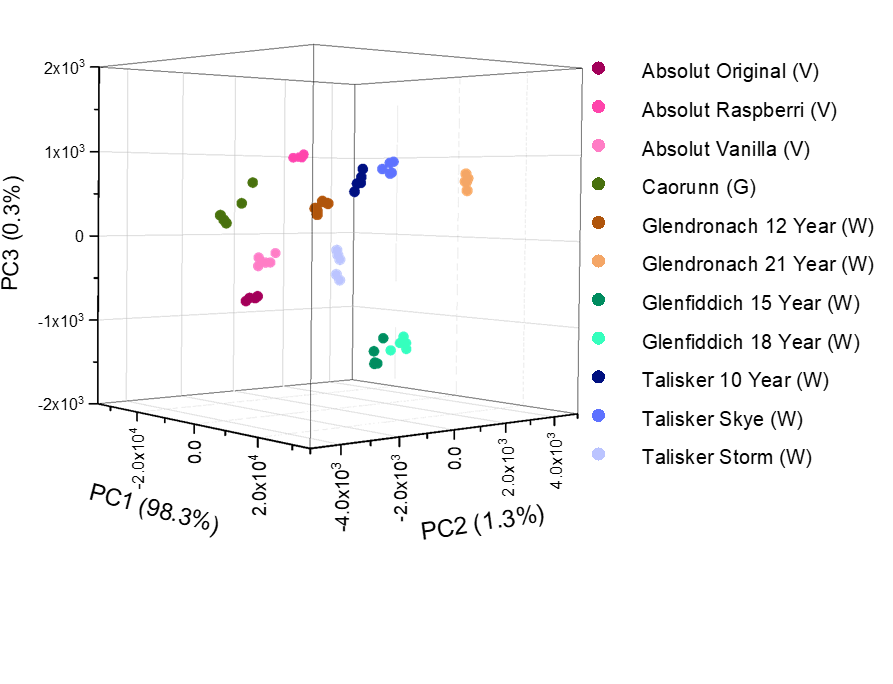}
\caption{Principal Component Analysis (PCA) scores plots discriminating between several spirits in their own commercial bottles. Samples were positioned at 25 mm from L\textsubscript{2} (Figure~\ref{scheme}). (V) denotes vodka, (G) denotes gin, (W) denotes whisky.}
\label{PCA}
\end{figure}

While conventional SORS configurations may be employed to probe the contents of bottles, a major disadvantage of this method is that the signal intensity decreases with an increasing offset distance as the excitation beam is typically focused at the surface or somewhere inside the sample while the collection is away from the position with the strongest excitation (focus). This makes the original very weak Raman signals even weaker and may cause strong background noise and undesired Raman signals from the excitation points. Here, we show the ability to reduce the Raman signals from the surface while maintaining a strong Raman signal from the contents inside of the bottle, using modest laser power (100 mW). Samples consisted of commercially available spirits contained in their original clear glass bottles. 

The effect of sample position with respect to the distance from L\textsubscript{2} along the optical axis was investigated (Fig.~\ref{Irisbottle}). 
By increasing the distance between L$_2$ and the surface of the bottle by 2 mm intervals, the obtained Raman spectra shows that the glass signal at 1370 cm\textsuperscript{-1} is reduced significantly whilst maintaining the strong ethanol Raman signal. The glass signal is reduced considerably at a distance of 8 mm from L$_2$, which corresponds to an radial offset distance of 0.15 mm on the surface of the bottle.

The spectra obtained from the ring-illumination system and a SORS configuration were compared with respect to the offset distances. The offset distance for the axicon-based system was determined by the radius of the ring excitation as illuminated on to the surface of the glass bottle. Fig.~\ref{AxVsSORS} shows the ratios of the glass peak at 1370 cm\textsuperscript{-1} and the ethanol peak at 881 cm\textsuperscript{-1} calculated for offset distances between 0-0.75 mm (peaks indicated by dotted lines in fig.~\ref{AxVsSORS}C). It is clear from Fig.~\ref{AxVsSORS} that compared to a SORS configuration, the axicon system can achieve a significant reduction in the glass signal with minimal offset. 

Comparing the signal afforded between a conventional backscattering Raman setup and the axicon-based Raman setup, there is an obvious reduction in the glass signal at 1370 cm\textsuperscript{-1}(Fig.~\ref{AxVsCr}). 
Eleven commercially available spirit samples were analysed, taking Raman spectra through their original glass bottle, at distance of 25 mm from L\textsubscript{2}. 
Principal component analysis (PCA) was used to analyse Raman spectra plotting the spectral features with the most variance (principal components) against each other. 
Here, the first three PCs account for $>99\%$ of the total variance of our data, thus the first three PCs have been plotted in Fig.~\ref{PCA}. 
From the PCA plot, it was observed that each of the beverages could be separated successfully. 
The plot suggests there is grouping between some of the distilleries, such as Talisker and Glenfiddich whiskies, which are found in their respective clusters. Further, we included several drink varieties which exhibit little to no fluorescence characteristics: vodka and gin. Four non-fluorescent samples consisting of 3 vodkas from the same brand, and 1 gin sample were included in the analysis, each displaying distinct separation in the PCA plot (Fig.~\ref{PCA}).

\section{Conclusions}
We have demonstrated an axicon-based system to enable Raman spectroscopic analysis of the contents of a bottle, by creating an annular beam formed on the bottle surface and a Bessel beam inside the bottle. 
This significantly reduces the Raman signal contributions from the bottle. We have used this setup to demonstrate the analysis of several distilled spirit drinks contained in their original commercial glass bottles, finding that there is a significant reduction of glass signal intensity while maintaining a strong signal afforded from the contents of the container. Further, this method possesses the advantage of being a simple modification to a standard epi-fluorescence optical arrangement by the introduction of an axicon lens. 
As such, it is potentially very attractive as a simple and low cost alternative to SORS configurations, providing a way of non-destructive and non-contact detection to precisely analyse the liquid contents without opening the bottle.\\

\begin{acknowledgments}
We thank the UK Engineering and Physical Sciences Research Council for funding through grants EP/R004854/1 and EP/P030017/1.
\end{acknowledgments}

\bibliography{throughbottle}

\end{document}